# P-wave arrival picking and first-motion polarity determination with deep learning

**Zachary E. Ross[1], Men-Andrin Meier[1], and Egill Hauksson[1]**

[1] Seismological Laboratory, California Institute of Technology, Pasadena, CA 91125

Corresponding author: Zachary Ross (zross@gps.caltech.edu)

**Key Points:**

- We train and validate convolutional neural networks to pick P-wave arrival times and first-motion polarities on 19.4 million seismograms
- Arrival time picks are within 0.028 s of the analyst pick 75% of the time, and first-motions are classified with 95% precision
- The remarkable performance of the trained networks suggests they can perform as well, or better, than human experts

**Abstract**

Determining earthquake hypocenters and focal mechanisms requires precisely measured P-wave arrival times and first-motion polarities. Automated algorithms for estimating these quantities have been less accurate than estimates by human experts, which is problematic for processing large data volumes. Here, we train convolutional neural networks to measure both quantities, which learn directly from seismograms without the need for feature extraction. The networks are trained on 18.2 million manually picked seismograms for the southern California region. Through cross-validation on 1.2 million independent seismograms, the differences between the automated and manual picks have a standard deviation of 0.023 seconds. The polarities determined by the classifier have a precision of 95% when compared with analyst-determined polarities. We show that the classifier picks more polarities overall than the analysts, without sacrificing quality, resulting in almost double the number of focal mechanisms. The remarkable precision of the trained networks indicates that they can perform as well, or better, than expert seismologists.

**1 Introduction**

Observed phase arrival times and first-motion polarities of seismic P-waves are essential ingredients in determining hypocenters and focal mechanisms (e.g. Hardebeck & Shearer, 2002; Yang et al., 2012). Historically, these quantities were measured manually by human experts, but as seismic networks have grown worldwide, such tasks have been increasingly taken up by automated algorithms. In applications of real-time seismology, such as earthquake early warning (Allen & Kanamori, 2003; Heaton, 1985; Satriano et al., 2011), all inference is necessarily performed with automated algorithms. Automated procedures additionally provide consistency and repeatability, whereas manual analysis may change over time or between different analysts. Consistent and well-characterized phase arrival picks are critically important for travel-time based inversion schemes such as hypocenter determinations and the accuracy of observed phase arrival times can be the limiting factor for seismic tomography studies (Allam & Ben-Zion, 2012; Diehl et al., 2009; Di Stefano et al., 2006).

Beginning with the short-term average/long-term average (STA/LTA) algorithm (Allen, 1982), a variety of classes of picking algorithms have been proposed, which include autoregressive methods (Sleeman & van Eck, 1999), higher-order statistics (Baillard et al., 2014; Ross et al., 2016; Saragiotis et al., 2002), predominant period (Hildyard et al., 2008), envelope functions (Baer & Kradolfer, 1987), and neural networks (Gentili & Michelini, 2006; Wang & Teng, 1997). Methods for picking the first-motion polarity include searching for zero crossings around the P-wave pick (Chen & Holland, 2016) and Bayesian inference schemes (Pugh et al., 2016). While there has been much success in the development and application of automated algorithms in seismology, they primarily are less precise than if a human performed the same task. This is likely because human analysts can simultaneously recognize a variety of general characteristics of an object (in this case, the appearance of an earthquake seismogram), while most automated algorithms aim only at a small number of characteristics that are formalized with simple threshold criteria (e.g. an amplitude threshold).

Machine learning and data mining algorithms provide an opportunity to significantly improve the performance of automated tasks in seismology because they allow for more complex inference approaches that mimic the behavior of the human mind (LeCun et al., 2015). Automated earthquake phase detection has been improved dramatically with template-waveform-based methods that search continuous seismic data for signals that are similar to previously detected ones (Shelly et al., 2013; Yoon et al., 2015; Skoumal et al., 2015; Ross et al., 2017). Classification of different types of seismic signals has been performed using Hidden Markov Models and neural networks (Hammer et al., 2012, and references thereisn; Mousavi et al., 2016), using both supervised and unsupervised approaches. Chen (2018) proposed an unsupervised microseismic picking algorithm that utilizes fuzzy clustering to identify signal onsets.

When sufficient amounts of labeled training data are available, supervised approaches enable direct quantification of the precision of a learning algorithm since the ground truth is known beforehand. Because of the decades long efforts of the Southern California Seismic Network (and its predecessors) to measure precise arrival times and first motion polarities by hand on a routine basis, we can take advantage of the rapid recent advances in neural network (NN) technology. In general, neural networks (NN) form a non-linear mapping function with a large number of terms (up to several millions for deep NN) that take a set of input values (e.g. the amplitudes of a seismogram or engineered features) and map them to a desired output (Figure 1). This output can be the prediction of a continuous variable (e.g. a phase onset time) or a class prediction (e.g. whether the first motion is up or down). The mapping function is organized in sequential layers of neurons, each of which is a simple function acts on incoming data and passes the result on to the next layer. The coefficients of the terms of the mapping function are empirically optimized with large amounts of data, such that a given set of input values leads to an output that is as close as possible to the desired output (e.g. maximally precise picks or correct class predications across a large validation data set).

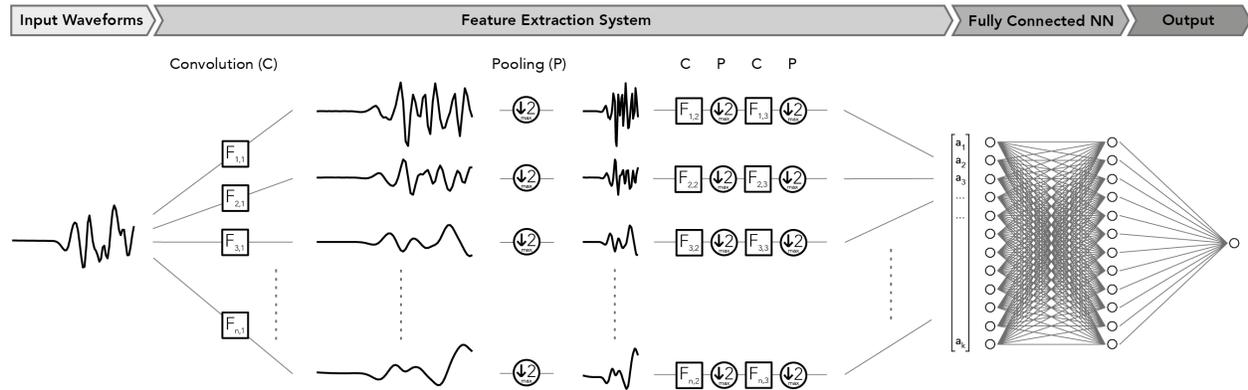

**Figure 1.** Cartoon depicting a CNN workflow for arrival time picking. In the first convolution step, the input seismogram is filtered in parallel with n=32 different filters with a length of 21 samples each. The filter specifications themselves are learned during the model training. The output of each filter is down-sampled in a subsequent pooling step by retaining the maximum of any two neighboring samples ('max pooling'). The process is repeated several times more, after which the output signals are concatenated and used as the input for a fully connected (FC) neural network. The convolutional network is a learnable feature extraction system that works together with a fully connected network for classification and regression tasks.

In recent years, the field of computer vision has undergone rapid transformation due to the emergence of convolutional neural networks (CNN) (LeCun et al., 1998; Krizhevsky et al., 2012; Simonyan & Zisserman, 2014; Szegedy et al., 2015; Zeiler & Fergus, 2014), a powerful variant of supervised machine learning. CNN differ from classical fully-connected NN in that they first use a set of locally-connected convolution and pooling layers that are fed with the data directly (Krizhevsky et al., 2012), rather than features extracted from the data. Each convolution layer consists of a set of learnable filters that are convolved with the outputs of a previous layer to identify patterns of interest anywhere within that data subset. In this case, learnable means that the coefficients of the filters are optimized along with all the other coefficients of the network during the training process. After convolution, pooling layers are commonly used to decimate the convolution output so that subsequent layers learn attributes of a re-scaled representation of the original input data. This helps recognizing variants of the same objects with different sizes, and it leads to an indirect connection between the only locally connected neurons of any individual pair of layers with the distant neurons of more shallow layers.

This first part of a CNN can be thought of as a feature extraction system that distills the relevant information from the input data, and then passes it on to a standard fully-connected NN. CNN are now the state-of-the-art approach in object detection and localization (Girshick et al., 2014; Krizhevsky et al., 2012). By their design they excel at performing pattern recognition that is invariant with respect to translation, scaling, and other types of distortions, which is a weakness of standard NN and other types of machine learning algorithms. As earthquake seismograms can be viewed as 1-dimensional images with three components, we demonstrate that the power of CNN algorithms can be readily and effectively applied to seismology problems.

Here we develop and apply a framework for automated P-wave picking and first-motion classification using two separate CNN. We utilize millions of picks and polarities determined by human experts at the Southern California Seismic Network (SCSN) from 2000-2017 to train and validate both networks. We show that the CNN produce extraordinarily precise picks and first motion polarity classifications that are comparable to, or better than, those made by humans.

**2 Data**

We used the records of 273,882 earthquakes recorded by the SCSN (Southern California Earthquake Data Center, 2013) from 2000-2017 at 692 stations. Seismograms were only used for stations within an epicentral distance of 120 km. The data are a mixture of HHZ, HNZ, and EHZ channels. These seismograms are associated with 4,847,248 manually determined P-wave picks, and 2,530,857 first-motion polarities assigned by SCSN analysts.

**3 Methods and Results**

To use CNN for picking the polarity and arrival time of a P-wave, we break the task into two primary steps. First, the onset of the P-wave needs to be located reliably, and second, the sign of the first swing of the P-wave needs to be determined. Our methodology uses a separate CNN for each task, in a manner that is very similar to object localization and detection within 2D images (Sermanet et al., 2013). First, a precise onset time is determined for each P-wave arrival using a CNN acting as a regressor, and second, the first-motion polarities of the seismograms are determined using a different CNN acting as a classifier.

### 3.1 P-wave picking

Before a CNN can be trained to pick P-waves, the waveform data must undergo pre-processing. All data are down-sampled to 100 Hz, detrended, and filtered with a causal Butterworth filter between 1-20 Hz. We only use the vertical component of the sensor on which the analyst picked the arrival times. The instrument response is not removed. Most SCSN stations have co-located sensors, and therefore this distinction is important to ensure that the algorithm training and testing is performed on the same data set that was used by the analysts. We randomly split the 4,847,248 records into a training set and a test (verification) set. Rather than split the data evenly between the two sets, we chose to have the training set consist of 75% of the records, with the remaining 25% forming the validation set. This allows more records to be used in the learning process, which benefits from having as much data as possible.

We then select a 4 s long feature window centered on the P-wave arrival. In order to mimic the situation where the true pick time is unknown, we perturb the center of the feature window with a uniform random perturbation shift between -0.5 s and 0.5 s. In the training data set we use each seismogram five times with different random windows, which artificially expands the training dataset to nearly 18.2 million records (e.g. Sermanet et al., 2013). The 400 waveform amplitude samples of each seismogram are the features that we use as input data. The maximum perturbation of 0.5 s ensures that enough of the P-wave is always inside the feature window. Example waveforms are shown in Figure S3. For future waveforms to be picked with an undetermined arrival time, the window length is large enough to use with theoretical arrival times computed from a 1D velocity model for selecting a window center. Next, the amplitudes in each feature window are normalized by the peak absolute amplitude in the window. This helps to suppress the influence of amplitude variations with magnitude, distance, and other factors, and is motivated by the bounded grayscale range [0, 1] used in image recognition algorithms (although here the range is [-1, 1]).

The CNN is then trained as a regressor using the randomly located P-wave arrival time within the feature window as the dependent variable. The CNN model used for the training process is summarized in Table 1. The convolution and full-connected layers use rectified linear units (ReLUs) as the activation function (Nair & Hinton, 2010), while the output layer uses linear activation. The training is performed using the Adam stochastic optimization algorithm (Kingma & Ba, 2014) using the default learning rate of 0.001, in batches of 480 seismograms. The batch size controls how many records used for each iteration of the learning process, while the learning rate is a hyperparameter of the optimization algorithm. We use a Huber loss function (Huber, 1964), and regularization is provided through batch normalization applied to all of the layers (Ioffe & Szegedy, 2015). We terminated the learning process when the validation loss had not decreased over the previous 5 epochs, and selected the model with the best results over the full training history. We varied the learning rate over the range of 0.0001-0.1 and found that the recommended value of 0.001 produced the best results when validated against the test data set. The learning process lasted for a total of 25 epochs, and Figure S1 shows the training and validation loss as a function of epoch number. Three NVIDIA GTX 1060 GPUs were utilized for training the model, and each epoch took approximately 15 minutes to complete.

Manuscript submitted to *J. Geophys. Res.-Solid Earth*

**Table 1.** Model architecture for picking P-wave arrival times. CBP = Convolution, batch normalization, pooling. FB = fully connected, batch normalization. F = fully connected.

| Layer | 1 | 2 | 3 | 4 | 5 | 6 |
|---|---|---|---|---|---|---|
| Stage | CBP | CBP | CBP | FB | FB | F |
| # Channels | 32 | 64 | 128 | 512 | 512 | 1 |
| Filter size | 21 | 15 | 11 | - | - | - |

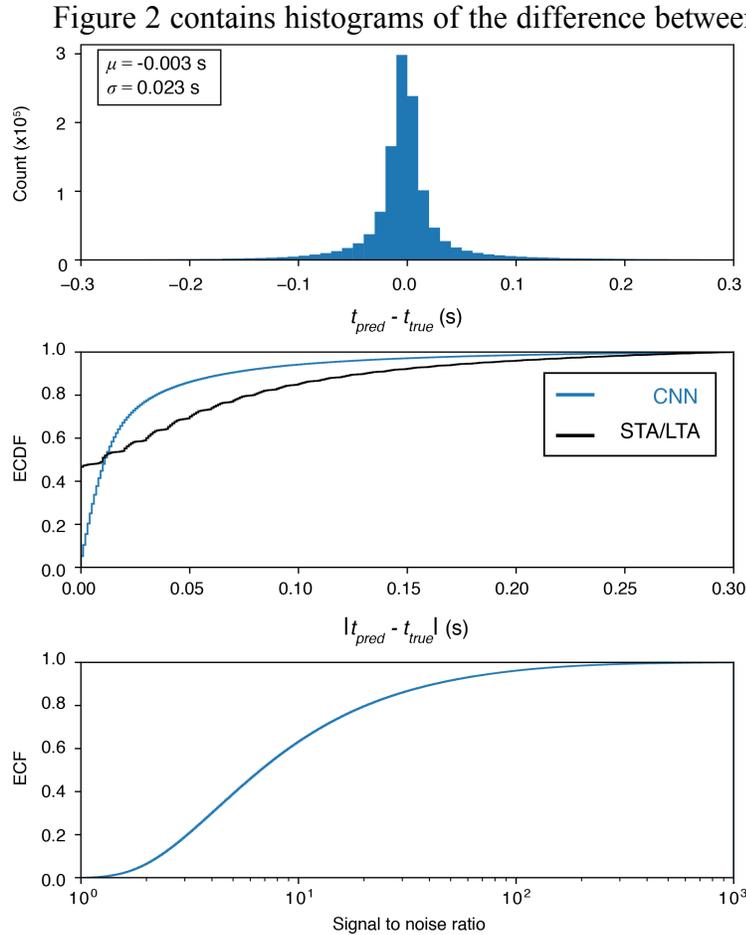

**Figure 2.** Summary statistics for the test dataset. 6.1 million records were used for validation. a) histogram of picking errors relative to analyst pick. b) cumulative histogram of the absolute picking error. c) cumulative histogram of SNR values for all P-wave onsets.

Figure 2 contains histograms of the difference between the predicted and analyst pick for the 6.1 million validation samples. The mean difference is $-3 \times 10^{-3}$ s, i.e. an order of magnitude smaller than the sampling interval of 0.01 s. The standard deviation is 0.023 s (Fig. 2a), and 75% of the picks are within 0.028 s of the analyst pick (Fig. 2b). The 90th percentile for the absolute differences is 0.074 s. While calculating the mean and standard deviation, the outer fence method was used to remove a very small number of extreme outliers, but for all other statistical metrics, these outliers were left in place. For comparison, picking errors from methods based on standard fully-connected neural networks have been reported with a standard deviation of 0.06-0.07 s (Gentili & Michelini, 2006), while errors from STA/LTA and kurtosis detectors generally have a standard deviation of 0.08-0.20 s (Gentili & Michelini, 2006; Nippress et al., 2010). Thus, the picks made using the trained convolutional network are about 3-10 times more precise than most commonly applied methods. Furthermore, the precision of the convolutional network is comparable to the errors believed present in the analyst picks (due to the presence of noise), and are based on all picks in the test dataset, without a single record being excluded. This is rather remarkable since most records have low signal/noise ratios (Figure 2c), which complicates the task of estimating phase arrival times. Examples of randomly chosen seismograms and the automated picks are shown in Figure S4.

Figure 3 displays percentiles of the absolute pick error as a function of signal-to-noise ratio (SNR), epicentral distance, and event magnitude. Here, SNR is defined as the ratio between the peak absolute amplitude in the 0.5s before and after the analyst P-wave pick in order to quantify the sharpness of the onset. There is a visible trend of decreasing pick error with SNR. The pick error increases rapidly for SNR < 5, which reflects the fact that true onsets under these conditions are likely to be beneath the noise level and therefore undefined. Pick error increases weakly with distance, presumably because SNR decreases with distance. The error is generally constant with magnitude. Overall, these numbers demonstrate the robustness of the method for picking P-wave arrivals with high precision across the entire data set.

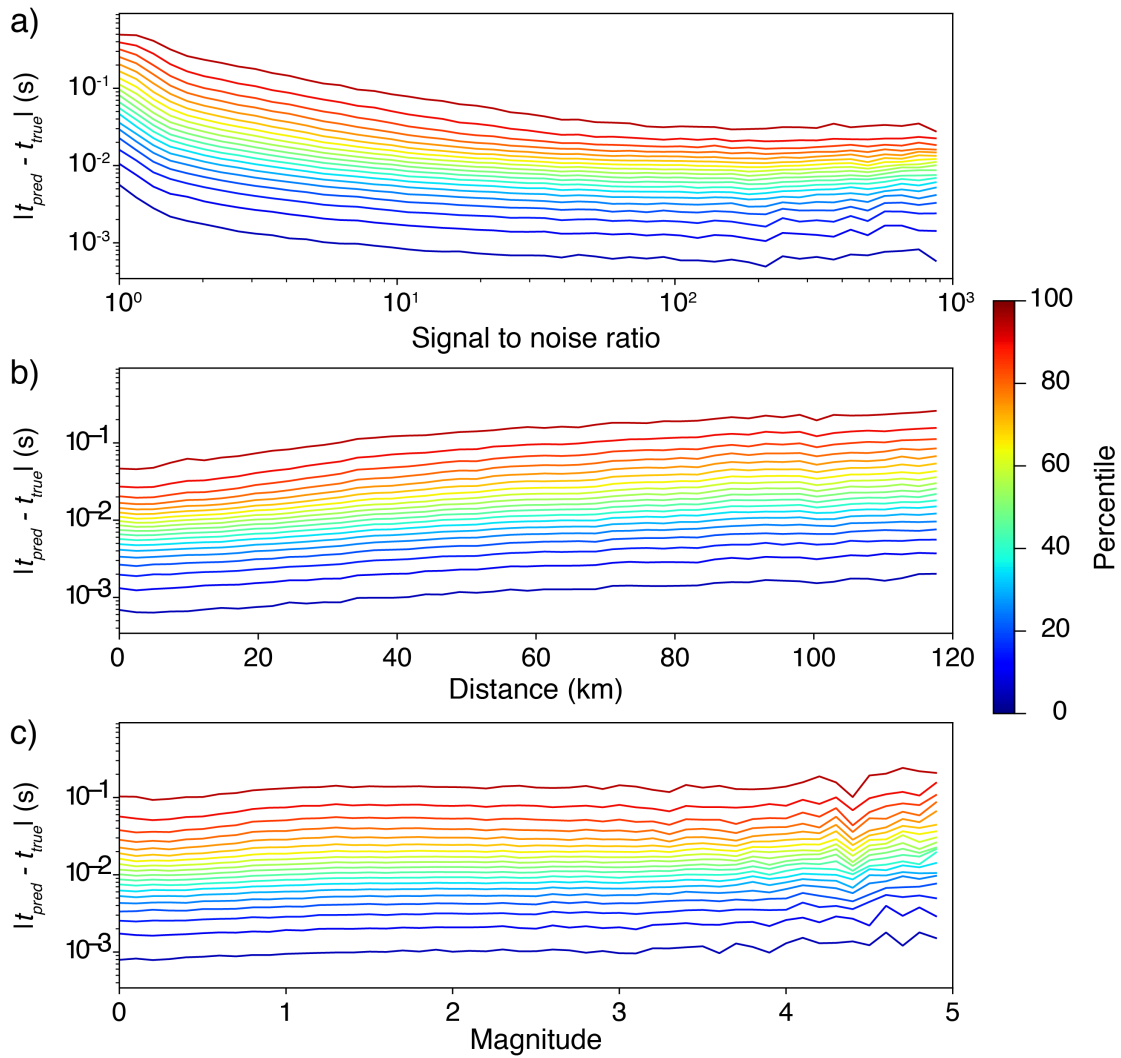

**Figure 3.** Distribution of absolute picking error as a function of SNR (a), epicentral distance (b) and magnitude (c). For records with low SNR, the network still mimics the human analysts' behavior with high precision.

### 3.2 First-motion polarity classification

With the ability to precisely pick the P-wave onset time, we now focus on the task of classifying first-motion polarities. The same pre-processing steps used for the arrival picking CNN are applied here as well. Each record is then labeled as up (u), down (d), or unknown (k) based on whether a first-motion polarity was assigned by the analyst for the station of interest. Records for which an analyst has determined a P-wave arrival time but not assigned a first-motion polarity are assigned label (k). This labeling process results in 2,525,947 records labeled up or down, and 2,321,301 records labeled unknown. These labels are not equally represented in the dataset, which can have an influence on the trained model. Therefore, we first identify the label with the fewest values of the three (d). This value (831,398) is then used for splitting the data, rather than from the total number, to ensure even representation of the three classes during the training process. The total number of seismograms for each class and data set are summarized in Table 2, including the augmented volume of seismograms for the training data.

**Table 2**. Number of seismograms for each class in the training and test sets.

| Class | Training | Test |
|---|---|---|
| Up | 4,156,990 | 586,018 |
| Down | 4,156,990 | 277,133 |
| Unknown | 4,156,990 | 1,489,903 |

These labeled samples are then used to train the second CNN with the same architecture as the pick time CNN, except that for this network the final output layer is replaced with a softmax activation function (classifier). Table 3 contains a diagram summarizing each of the layers and the output classification scheme. Softmax is a multi-dimensional generalization of the logistic function and is widely used with neural networks to map a set of input values into a set of output values in the range [0, 1], such that the outputs sum to 1. We again train this CNN using the Adam optimization algorithm, but using a cross-entropy loss function. A learning rate of 0.001 is used to train the CNN in batches of 480, for 8 epochs, with a patience value of 5.

**Table 3.** Model architecture for determining first motion polarities. CBP = Convolution, batch normalization, pooling. FB = fully connected, batch normalization. F = fully connected.

| Layer | 1 | 2 | 3 | 4 | 5 | 6 |
|---|---|---|---|---|---|---|
| Stage | CBP | CBP | CBP | FB | FB | F |
| # Channels | 32 | 64 | 128 | 512 | 512 | 3 |
| Filter size | 21 | 15 | 11 | - | - | - |

We now evaluate the success of the CNN to classify the first motion polarities for all records of the validation data set (Figure 4). The precision for determining a given class is defined as the number of true positives divided by the total number of records assigned to the class by the CNN. Here we also have the possibility of making picks (correctly) that an analyst could not. However, because we cannot validate these classifications without ground truth, we do not include them in this particular evaluation (they will be examined in detail subsequently). As an example, for class (u), precision is defined as the number of cases when both the CNN and analysts assigned (u) (true positives), divided and the total number of number of cases when the CNN assigned (u), and the analysts assigned either (u) or (d) (true positives plus false positives). The overall precision for class (u) is 0.97, while for (d) it is 0.93. This means that 3% and 7% of the picks labeled as (u) or (d), respectively, by the CNN were assigned the opposite first-motion polarity by the analysts. Recall, which measures the fraction of cases of a given class (made by the analysts) that were correctly identified by the classifier, is 0.80 for up and 0.81 for down. These numbers indicate that about 20% of the polarity picks made by the analysts were assigned (k) by the CNN. However, as we will demonstrate subsequently, the CNN makes 236,237 (~27%) additional picks that the analysts did not, and their good agreement with independently determined focal mechanisms suggests that these polarities are typically accurate as well.

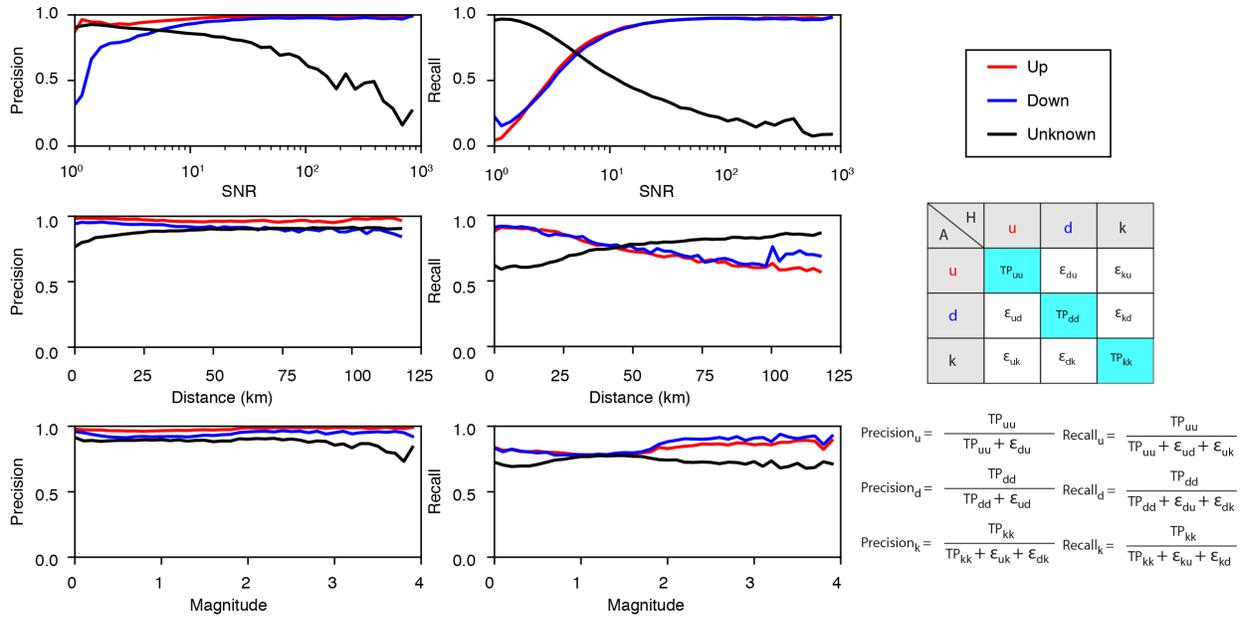

**Figure 4.** First-motion polarity classification results. Precision and recall as a function of SNR, epicentral distance, and magnitude. Up and down first-motion picks have an average precision of more than 95%.

These precision and recall numbers represent averages for the entire data set. In Figure 4a we show the average precision and recall of the CNN in a range of SNR bins. When the method chooses to assign a first-motion polarity (i.e. to assign a label other than 'k'), the precision is near 98% for SNR > 10, and only marginally worse for low SNR conditions. The recall generally increases with SNR, reaching around 87% for SNR 10. The fact that the recall is lower than the precision suggests that, if the classification is uncertain due to low a SNR, the network acts conservatively in that it assigns label (k), rather than (u) or (d), thereby decreasing recall rather than precision. Figure 4b shows the precision and recall of the CNN as a function of epicentral distance. While precision generally varies little with distance, recall decreases steadily with distance, presumably because the lower SNR leads to more assignments of label (k). Figure 4c shows the precision and recall of the CNN as a function of magnitude. The precision is again about 95% for the full range of magnitudes, but the recall slowly increases with magnitude.

## 3.3 Focal mechanism comparisons

The first-motion polarities are used to determine earthquake focal mechanisms. We use the all of the predicted first-motions from the validation data set to calculate focal mechanisms with the HASH method (Hardebeck & Shearer, 2002) to invert the polarities. The velocity models provided with the HASH code for southern California are used. The minimum number of required polarities is 8, and the maximum azimuthal and takeoff angle gaps allowed are 90 and 60 degrees, respectively. In this study, we specifically do not use S/P amplitude ratios to constrain the focal mechanisms in order to allow direct evaluation of the quality and number of mechanisms by adding first motions from the CNN.

HASH determines focal mechanism quality based on uncertainty in the fault plane, and assigns quality labels of A-F. We compare our results to focal mechanisms produced from manually determined polarities made by SCSN analysts, using all mechanisms with A, B, C, and D quality. Of the 148,439 events in the test data set, 4613 events in the SCSN FM catalog had quality A-D, whereas 6003 events in the CNN dataset did. For all four quality classes the CNN dataset reached higher numbers of events. Specifically, the number of events with A, B, C, and D qualities increased by 84%, 90%, 51%, and 18%, respectively. Thus, the highest quality grades increased by the highest percentage. These results demonstrate that the CNN classifier determines first motion polarities in such a reliable manner that the resulting catalog contains almost double the number of high quality FMs as a catalog based on human polarity determinations.

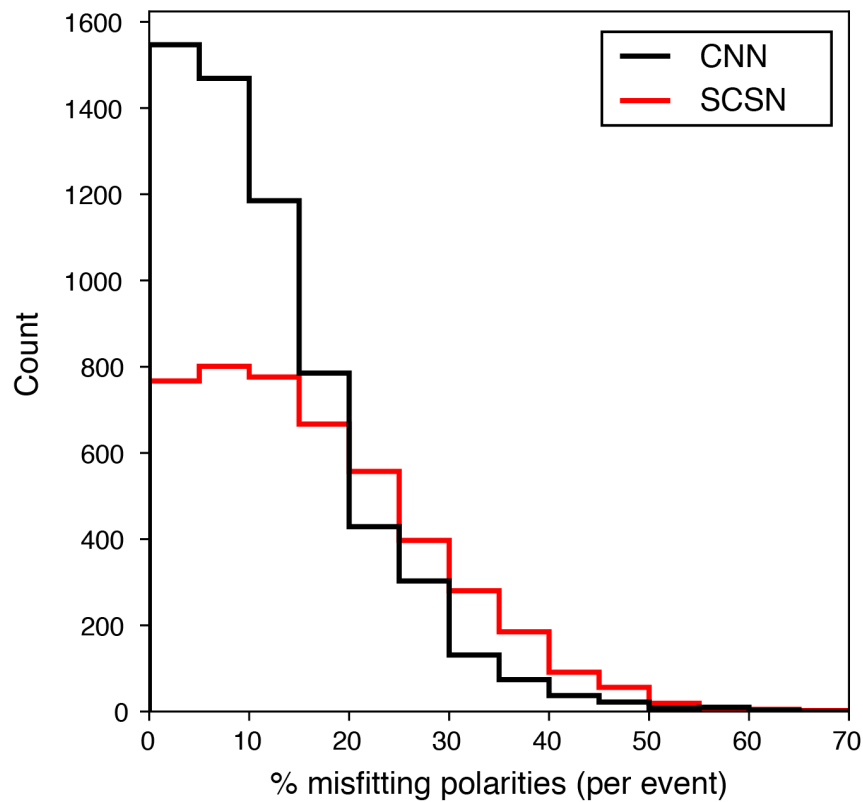

**Figure 5**. Comparison of focal mechanism quality between automated (CNN) and manual datasets (SCSN). The percentage of misfitting polarities is defined per event using the best-fitting focal mechanism. The CNN leads to more focal mechanisms determined overall, and they are of higher quality than those from the manually picked data.

To further investigate the quality of the focal mechanism catalogs, we examine the percentage of misfitting polarities for the best-fitting focal mechanism of each event. Here, a value of 0% indicates that all the polarities are the same sign as the theoretical radiation pattern of the focal mechanism. Figure 5 contains histograms of this metric for all events, for both the SCSN catalog (red) and the CNN catalog (black). The CNN has significantly fewer values with misfit percentages greater than 30%, and nearly twice as many values where misfit percentages are smaller than 10%, relative to the SCSN catalog. Since the same three parameter model (strike, dip, rake) is fit to both datasets for the same events, these results indicate that the CNN is able to make more picks than humans can, with even greater precision.

### 3.4 Using the trained networks for future processing

The methodology as described insofar has focused on the details for training convolutional networks as well as the performance as applied to specific validation records. Here we discuss how the trained networks might be utilized in practice. These networks were not designed to detect earthquakes; rather they were optimized to precisely measure specific P-wave attributes assuming that the earthquake has already been detected beforehand. Thus, some method must be utilized to reliably detect earthquakes in the continuous data as an initial step. After that, the data must be filtered between 1-20 Hz and resampled (if necessary) to 100 Hz.

Since the networks require a 4 s window of data as input, this window must be chosen somehow. One simple solution is to use 1D travel time predictions to define the center of the feature window, with the SNR being checked to ensure the window was not only noise. Alternatively, an STA/LTA detector could be used with a simple trigger threshold (e.g. 5.0) to define the window center. This issue is primarily for the arrival picking CNN, as it always returns a value regardless of whether any signal is present in the window. The first-motion classifier, however, was designed to label windows as undefined if the SNR is too low, and therefore is generally not susceptible to these issues. In a real-time seismic processing system, these networks could be triggered once an earthquake has been detected. CNN architectures could also be used for earthquake detection (Perol et al., 2018).

Training a convolutional network is computationally demanding, and in this study, we utilized 3 NVIDIA GTX 1060 GPUs to accomplish this. GPUs are well-suited for the massively parallel floating point arithmetic that deep learning requires; however for forward prediction and classification of individual seismograms, a multicore CPU may still work sufficiently depending on the amount of data to be processed. Modern GPUs have sizable memory, which can enable large numbers of seismograms to be processed simultaneously with limited transfer between the CPU and GPU. The performance of the algorithms, however, is highly dependent on the hardware itself, with GPU technology rapidly improving by the year.

### 4 Discussion

Phase arrival times are generally the first type of information determined about an earthquake, and form the basis for a wide range of subsequent seismological measurements. Improvements in the methods used for measuring these quantities can propagate into every subsequent measurement, including locations, magnitudes, and source properties. The typical approach to automated phase picking in seismology has been to calculate characteristic functions that are likely to indicate a phase arrival. Here, we instead treat the problem as one of image recognition, in which CNN were trained to learn the general characteristics of P-wave onsets. We forgo the process of feature extraction, and instead use the seismogram directly as an input with only minor pre-processing. This enables the problem to be solved in a manner analogous to how a human would solve it.

When working with CNN, there are many choices to be made for parameters and algorithms involved in the training process. Some of the parameters were chosen from trial and error, while others were more rigorously tested over a specific range. The window length of the seismograms was chosen to be 4 sec to enable the window center to be defined using predicted arrival times from a 1D model, while being tolerant of velocity model uncertainty. However, values in the range 2-6 seconds also lead to similar results. For the model parameters, we explored various configurations and settled on the final model because it provided the best results against the analyst-picked data. CNN model construction is presently a very active subject of research, and designs are generally chosen through experimentation, since the ground truth can be used for direct validation of a model architecture. Other individuals may use the same training data and find a better model in the future. To this end we provide the full training and validation dataset as a single compact hdf5 file so that anyone interested in improving upon our CNN design can use the exact same data for training and testing. Having a standardized seismological data set will ensure that the results of future algorithms proposed are translatable between studies.

The choices necessary for designing a CNN architecture, however, only determine the general problem set up, and represent rather soft constraints on how the regression and classification is performed. The more traditional automated picking methods, on the other hand, typically involve hard thresholds, e.g. a trigger threshold for the characteristic function, and these are often difficult to optimize in a systematic way. The ability of CNN to work directly with seismograms to systematically optimize decision boundaries is one of the main reasons why such approaches can outperform the traditional methods.

Phase picking and first-motion classification on noisy waveform data are often difficult tasks even for humans to perform, and therefore handmade picks and first-motions are also imperfect. It is reasonable to assume that handmade P-wave picks are probably only accurate to within a few samples for most seismograms, depending heavily on the signal to noise ratio at the onset itself. This is because the true onset for many seismograms may in fact be beneath the noise level, and therefore only known to within some subjective uncertainty range. Since the P-wave picks made by the CNN are within this same uncertainty range, this means that the picks are indistinguishable in quality from those made by human beings at a statistical level. There is some evidence that handmade first-motion polarities are only correct about 80-90% of the time (Hardebeck & Shearer, 2002), likely also resulting from the first swing of the P-wave actually being below the noise level. Incorrect assignments of the manual first-motion polarities would result in the precision of the CNN being lowered in our tests. The first-motion polarities determined by the CNN have precision and recall of about 95% and 80%, respectively, but the method made 30% more measurements that the analysts could not. Since there is no ground truth for these extra cases, we evaluated the overall quality of the CNN determined dataset against the analyst determined dataset by inverting for focal mechanisms. As with the P-wave picks, this analysis suggest that the first-motion polarities determined by the CNN are indistinguishable in quality from what a human expert can do, and are arguably even more accurate.

In this study, we trained two CNN from millions of records that were laboriously hand-picked and labeled over a period of almost two decades. Since the training process results in the CNN learning how to make the same decisions that the SCSN analysts made, these networks can therefore be viewed as containing the full knowledge of the data archives inside of them. Thus, each future automated pick will draw on the collective experience of the analysts at the network, which can only improve with more data in the future.

**Acknowledgments**

This study was performed using TensorFlow. Data used in this study were collected by the Caltech/USGS Southern California Seismic Network. doi:10.7914/SN/CI; and distributed by the Southern California Earthquake Center; doi:10.7909/C3WD3xH1. The study was supported by the Gordon and Betty Moore Foundation and NSF award EAR-1550704. There are no real or perceived financial conflicts of interest for any author.